\documentclass[final,1p,times]{elsarticle}

\usepackage{graphicx}
\usepackage{amssymb}
\usepackage{amsmath}
\usepackage{acronym}        
\usepackage{xspace}         
\usepackage[hyphens]{url}   
\usepackage{hyperref}       
\usepackage{cleveref}       
\usepackage{siunitx}        
\usepackage{booktabs}       
\usepackage{hepnames}       

\DeclareSIUnit[number-unit-product = \;]{\permil}{\textperthousand}
\DeclareSIUnit[number-unit-product = \;]\byte{B}
\DeclareSIUnit[number-unit-product = \;]\flops{FLOPS}
\DeclareBinaryPrefix\kibi{Ki}{10}
\DeclareBinaryPrefix\mibi{Mi}{20}
\DeclareBinaryPrefix\gibi{Gi}{20}

\hypersetup{%
    colorlinks,
    citecolor=blue,
    filecolor=black,
    linkcolor=black,
    urlcolor=black
}

\newcommand\picheight{.66\textwidth}
\newcommand\picpath{plots}

\newcommand{\T}[1]{\mathrm{#1}}

\newcommand{\E}{\T e}
\newcommand{\I}{\T i}

\newcommand{\R}{\mathbb{R}}

\newcommand\no{\notag\\}
\newcommand\po{\;.}
\newcommand\co{\;,}

\newcommand\Rcite[1]{Ref.~\cite{#1}}

\newcommand\ie{i.e.}
\newcommand\eg{e.g.}
\newcommand\cf{cf.}
\newcommand\wrt{with respect to\xspace}

\crefformat{section}{Section~#2#1#3}
\crefformat{figure}{Figure~#2#1#3}
\crefformat{table}{Table~#2#1#3}
\crefformat{chapter}{Chapter~#2#1#3}
\crefformat{appendix}{Appendix~#2#1#3}

\newcommand\lang[1]{\texttt{#1}}
\newcommand\prog[1]{\textsc{#1}}
\newcommand\code[1]{\texttt{#1}}
\newcommand\cd[1]{\code{#1}}
\newcommand\hardware[1]{{#1}}
\newcommand\order[1]{\ensuremath{\mathcal{O}\left(#1\right)}}

\newcommand\OMEGA{\prog{O'Mega}\xspace}
\newcommand\FORM{\prog{FORM}\xspace}
\newcommand\highperformance{high-performance\xspace}
\newcommand\HighPerformance{High-Performance\xspace}
\newcommand\subinstructions{sub-\-in\-struct\-ions\xspace}
\newcommand\Subinstructions{Sub-\-in\-struct\-ions\xspace}

\newcommand\Bytecode{Byte code\xspace}
\newcommand\bytecode{byte code\xspace}
\newcommand\bytecodes{byte codes\xspace}

\newacro{1POW}{one-particle off-shell wave function}
\newacro{APS}{antenna pole structure}
\newacro{BSM}{beyond the SM}
\newacro{CERN}{European Organization for Nuclear Research}
\newacro{CMF}{center of mass frame}
\newacro{CPU}{central processing unit}
\newacro{CSE}{commmon subexpression elimination}
\newacro{FPGA}{Field programmable gate array}
\newacro{GPU}{graphics processing unit}
\newacro{HL-LHC}{High Luminosity LHC}
\newacro{HMC}{Helicity Monte Carlo}
\newacro{ILC}{International Linear Collider}
\newacro{LCA}{leading color approximation}
\newacro{LHC}{Large Hadron Collider}
\newacro{LHS}{left-hand side}
\newacro{LO}{leading order}
\newacro{LSZ}{Lehmann-Symanzik-Zimmermann}
\newacro{MC}{Monte Carlo}
\newacro{MIC}{Many Integrated Cores}
\newacro{MSSM}{minimal supersymmetric SM}
\newacro{NLO}{next-to-leading order}
\newacro{NNLO}{next-to-next-to-leading order}
\newacro{NUMA}{Non-Uniform Memory Access}
\newacro{opcode}{operation code}
\newacro{OVM}{O'Mega virtual machine}
\newacro{PDG}{Particle Data Group}
\newacro{QCD}{quantum chromodynamics}
\newacro{QED}{quantum electrodynamics}
\newacro{RHS}{right-hand side}
\newacro{SIMD}{single instruction multiple data}
\newacro{SM}{standard model}
\newacro{VM}{virtual machine}

\biboptions{sort&compress}

\newcounter{bla}

\journal{Computer Physics Communications}

\begin{document}

\begin{frontmatter}
\begin{flushright}
  \normalsize{} DESY 14--206
\end{flushright}

\title{Simple, Parallel, \HighPerformance{} Virtual Machines for Extreme
Computations}

\author[desy,uniwue]{Bijan Chokoufe Nejad}
\ead{bijan.chokoufe@desy.de}
\author[uniwue]{Thorsten Ohl}
\ead{ohl@physik.uni-wuerzburg.de}
\author[desy]{J\"urgen Reuter}
\ead{juergen.reuter@desy.de}
\address[desy]{DESY Theory Group, Notkestr. 85, D-22607 Hamburg}
\address[uniwue]{%
  University of W\"urzburg, Emil-Hilb-Weg 22, 97074 W\"urzburg, Germany
}

\begin{abstract}
We introduce a \highperformance{} \ac{VM} written in a numerically fast language
like \lang{Fortran} or \lang{C} to evaluate very large expressions.
We discuss the general concept of how to perform computations in terms of a
\ac{VM} and present specifically a \ac{VM} that is able to compute tree-level
cross sections for any number of external legs, given the corresponding
\bytecode{} from the optimal matrix element generator, \OMEGA{}.
Furthermore, this approach allows to formulate the parallel computation of a
single phase space point in a simple and obvious way.
We analyze hereby the scaling behaviour with multiple threads as well as the
benefits and drawbacks that are introduced with this method.
Our implementation of a \ac{VM} can run faster than the corresponding native,
compiled code for certain processes and compilers, especially for very high
multiplicities,
and has in general runtimes in the same order of magnitude.
By avoiding the tedious compile and link steps, which may fail for source
code files of gigabyte sizes, new processes or complex higher order
corrections that are currently out of reach could be evaluated with a \ac{VM}
given enough computing power.
\end{abstract}


\end{frontmatter}

\section{Introduction}
\label{s:intro}
%
Computations in high energy physics tend to hit the limits of what is
computationally feasible.
Setting demanding grid computations aside, one encounters even in perturbative
calculations expressions of cross sections of enormous size.
Such computations for the \ac{LHC}, its upgrade the \ac{HL-LHC} or the planned
\ac{ILC} are and keep getting more challenging as cross sections are needed for
a high number of external particles and to increasing precision to match the
experimental efforts.
When facing such problems, a compromise has to be made, in order to have a
maintainable and extendible solution for the developer and at the same time fast
execution of the code.
The latter cannot be overrated as the same code, typically representing a
certain process, has to be evaluated billions of times with different input
data for the Monte Carlo integration and/or parameter scans.

A popular approach to solve this problem is a me\-ta-\-pro\-gramm\-ing ansatz, \ie{}
to determine the expression of a cross section itself in a higher level
programming language like \lang{Mathematica}, \lang{OCaml},
\lang{FORM} or \lang{Python} while the numerical evaluation is performed in
\highperformance{} languages like \lang{Fortran} or \lang{C}.
Hereby, the expression is vastly reduced with computer algebra and tailored
algorithms on the higher level to make the execution later on as fast as
possible.
Examples for this are the tree-level and one-loop matrix element generators
\prog{MadGraph}~\cite{MadGraph2011},
\prog{FormCalc}~\cite{FormCalc1998,FormCalc2013} or \OMEGA~\cite{Omega2001}.
A problem, however, arises when the expression becomes so large that it is
impossible to compile and link, and hence to evaluate numerically, due to the
sheer size.
In \lang{Fortran}, which is known for its excellent numerical performance, we
typically encounter this problem for source code of gigabyte sizes irrespective
of the available memory.
This problem is also being addressed by the project
\prog{HepGame}~\cite{HEPGAME2014} that is based on
\lang{Form}~\cite{Form2012,Form2014} and aims to reduce the code size before
compilation by using new concepts from game theory like Monte Carlo tree
searches.
Furthermore, we should mention \prog{haggies}~\cite{haggies2010}, written in
Java, which also generates optimised programs for efficient numerical
evaluation of mathematical expressions using multivariate Horner-schemes and
\ac{CSE} to reduce the source code size.

In this paper, we show how to circumvent the tedious compile step in between
completely by using a \ac{VM}.
To avoid confusions, we have to define what we mean with the term \ac{VM}.
A \ac{VM} is in our context a compiled program, an interpreter, that is able
to read instructions, in the form of \emph{\bytecode}, from disk and perform an
arbitrary number of operations out of a finite \emph{instruction set}.
We do not refer to any sort of operating system emulation that is
commonly encountered under the term \ac{VM}.
Also the parallel virtual machine (PVM)~\cite{PVM} is a completely different
idea, combining a network of multiple computers to one \ac{VM}.
Far closer to our \ac{VM} is the \ac{VM} used in the open-source project
\prog{numexpr}~\cite{numexpr}.
Their \ac{VM} is written in \lang{C} and specializes on the fast numerical
expression evaluation of very large arrays in \lang{Python} by dividing array
operands in chunks that easily fit in the cache of the \ac{CPU} and avoiding the
creation of temporary arrays.
Though the idea is related, in our application we have comparably small arrays
per instruction and can hence not benefit from this project.
We want to stress that a \ac{VM} allows the complexity of the computation to
be only set by the available hardware and not limited by software design or
intermediate steps.
Furthermore, we will show that a \ac{VM} is easy to implement and makes
parallel evaluation obvious.

An important concern is of course whether the \ac{VM} can still compete with
compiled code in terms of speed.
The instructions have to be translated by the \ac{VM} to actual machine code,
which is a potential overhead.
However, in the computation of matrix elements there are typically a lot of
complex scalar products involved in a single instruction implying that this
overhead plays a minor role.
What we explicitly give up are the optimizations that the compiler can perform
in the context of multiple instructions like \ac{CSE} and data and instruction
prefetching.
Of these, at least the \ac{CSE} can be done beforehand
by constructing the \bytecode{} with the lowest number of common
subexpressions on the higher level.
In fact, we will show that a \ac{VM} can even be faster than compiled code for
certain processes and compilers since the \ac{VM} also benefits from the fact
that instruction cache misses are less likely in this formulation.

We will apply the concept of a \ac{VM} to the tree-level Optimizing Matrix
Element Generator, \OMEGA{}, to allow the computation of higher multiplicities of
colored particles given the same hardware.
This does obviously not imply that the presented computational method is
restricted to tree-level computations.
When trying to obtain higher order cross-sections the same problem can arise
even for less external particles, due to the inherent complexity of the
computation.
We expect that \ac{VM} implementations in such environments are a possible way
to go beyond what is nowadays considered as still feasible.

Apart from cross sections, we believe that the problem of evaluating huge
expressions numerically is a more general one, just as algebraic tools like
\FORM{}~\cite{Form2012,Form2014} or integration tools like
\prog{CUBA}~\cite{Cuba2005} are useful beyond their original field of study.
Therefore, we will tackle this problem at first in a rather general way in
\Cref{s:virtual_machine}, before we turn to the implementation of the \ac{OVM}
in \Cref{s:omega_virtual_machine}.
Then we benchmark this proposal in \Cref{s:benchmarks} and conclude with a
summary of our findings and a small technical outlook in \Cref{s:conclusions}.
\section{General Virtual Machines}
\label{s:virtual_machine}
We will describe in this section the necessary components to perform a
computation with a \ac{VM}.
The \bytecode{} plays a central role as it embodies all nontrivial information
about how to compute the object of desire.
One might imagine the \ac{VM} as a machine, which has a number of registers, and
is given instructions how to act on them.
This picture is quite similar to a \ac{CPU}, except that we are
doing this on a higher level, \ie{} our registers are arrays of \eg{} wave
functions or momenta and the instructions can encode scalar products or more
complicated expressions.
In~\ref{a:a_trivial_example}, we also give a purely mathematical example
implementation of a \ac{VM}.
This code is well suited for adaption to other problems as it has no
dependency on external libraries and still includes all of the necessary
infrastructure.
\subsection{Byte Code}
\label{ss:vm:byte_code}
For the dynamic construction of the \ac{VM}, it is necessary to include a
\emph{header} in the \bytecode{}, which contains the number of objects that have
to be allocated.
For convenience, it is also useful to have some version numbers that document
which physical or mathematical constants should be used together with this
\bytecode{} or comments to indicate how it was produced.
Optionally, one can add after the header tables of precomputed parameters, like
information about the involved helicities, color or flavor.
After this the body of instructions follows, whereby each line corresponds to a
certain operation that the \ac{VM} should perform on its registers.

We encode the instructions in pure integers inside a simple \lang{ASCII}
\bytecode{} such that it is in principle human-readable if the meanings of the
numbers are known.
Though the use of mere numbers does not exploit the full dictionary of
\code{ASCII},
it allows a very fast integer line-by-line scan of the \bytecode{} in the
\lang{Fortran} \ac{VM} and avoids a translation step because integers are
already addresses in arrays.
Since the initialization is, however, very fast compared to the runtime, this
could be optimized with a binary format, if the size of the \bytecode{} becomes
a problem.
As the \bytecode{} size is about a factor of ten smaller compared to the native
source code, as shown in \cref{ss:bytecode_generation} this is not yet a concern
for our application.
The fact that our \bytecode{} is portable and platform independent is a positive
surplus when calculations are performed on clusters.

The first number of an instruction is the \emph{\ac{opcode}} that specifies
which operation will be performed.
For illustration, consider the example
\begin{verbatim}
1 5 4 3
\end{verbatim}
which could be translated into
\code{momentum(5) = momentum(4) + momentum(3)}, a typical operation to compute
the $s$-channel momentum in a $2\,\to\,2$ scattering process.
Depending on the context, set by the \ac{opcode}, the following numbers have
different meanings but are typically addresses, \ie{} indices, of objects, or
specify how exactly the function should act on the operands, by what numbers the
result should be multiplied, etc.

When designing the \bytecode{} of a \highperformance{} \ac{VM}, the line length
should be chosen such that the most frequent operations fit within a line.
Complex operations that would increase the line length significantly above the
average requirement, can also be split in multiple lines by using
\emph{\subinstructions}, which are introduced in
\cref{ss:general_parallelization}.
\subsection{Interpreter}
\label{ss:vm:interpreter}
%
The interpreter is a very simple program that reads the \bytecode{} into memory
and then loops over the instruction block with a \code{decode} function, which
is basically a \code{select/case} statement depending on the \ac{opcode}.
The instructions can be instantly, compared to the execution time of the
relevant instructions, translated to physical machine code, since the different
types of operations are already compiled and only the memory locations of the
objects have to be inserted.
The compilation of the \ac{VM} itself is very fast and has only to be done once
which is handy for the use of many \bytecodes{} and necessary for extreme
computations as motivated above.

Two things have to be adapted in the interpreter of the \ac{VM}, when one wants
to tackle a new type of problem, \eg{} when going from tree-level to one-loop.
At first, one has to specify, where and with which types to expect header,
comments, tables and instructions\footnote{One could, in principle, also
  determine this dynamically by using a certain markup, if one has the desire to
  do so.}.
Furthermore, the \code{decode} function needs to be able to translate any
instruction line into operations on registers, \ie{} all \acp{opcode} have to be
implemented.
The functions can be arbitrarily complex and are also allowed to call external
libraries, though most likely better performance is achieved by keeping things
as simple as possible.
Especially, with parallelization  in mind, it is desirable to have roughly the
same amount of computation time for different instructions, to ensure an even
workload and hereby minimizing idle times at synchronization points.

Given this environment, the \bytecode{} file that is given to the \ac{VM}
completely dictates the specific problem, or process in the cross section
context, that should be computed.
Input data or external parameters are given as arguments to the function call of
the \ac{VM}.
The calling application has of course to make sure that these parameters match
the corresponding \bytecode{}, which can be ensured with version numbers.
\subsection{Parallelization}
\label{ss:general_parallelization}
%
The generation of events for collider physics usually parallelize trivially.
Since an integral is in most cases needed,
the same code is just evaluated multiple times with different input data.
The situation can change, however, for an extreme computation that already uses
all caches.
Depending on the size of the caches and the scheduler, evaluating such code
with multiple data at the same time, can run even slower than the
single-threaded execution.
Obviously, the computation is then so large, containing numerous objects, that
it is worth trying to parallelize the execution with a single set of input
data with shared memory.

Developing truly parallel code for a complicated calculation, however, is a
non-trivial task since race conditions have to be kept in mind at all times.
Furthermore, physicists have to delve into the frameworks like \lang{OpenMP} or
\lang{MPI} to find the best parallelization method for each piece, which is
time consuming and likely to introduce bugs that are hard to find.
The \bytecode{} gives us the opportunity to write the parallel computation in an
obvious fashion that is both easy to generate and to implement in the \ac{VM}.
The idea is to split the \bytecode{} into recursion \emph{levels}, whereby in
each level all \emph{building blocks} are non-nested and may be computed in
parallel.
Different levels are separated by necessary synchronization points at which
threads have to wait until intermediate results are communicated and which can
be represented in the \bytecode{} with a zero \ac{opcode}.
It is clear that one should aim to keep the number of synchronization points to
the inherent minimum of the computation for optimal performance.

\begin{figure}[htbp]
\centering
\includegraphics[width=.9\columnwidth,height=.5\textwidth,keepaspectratio]
{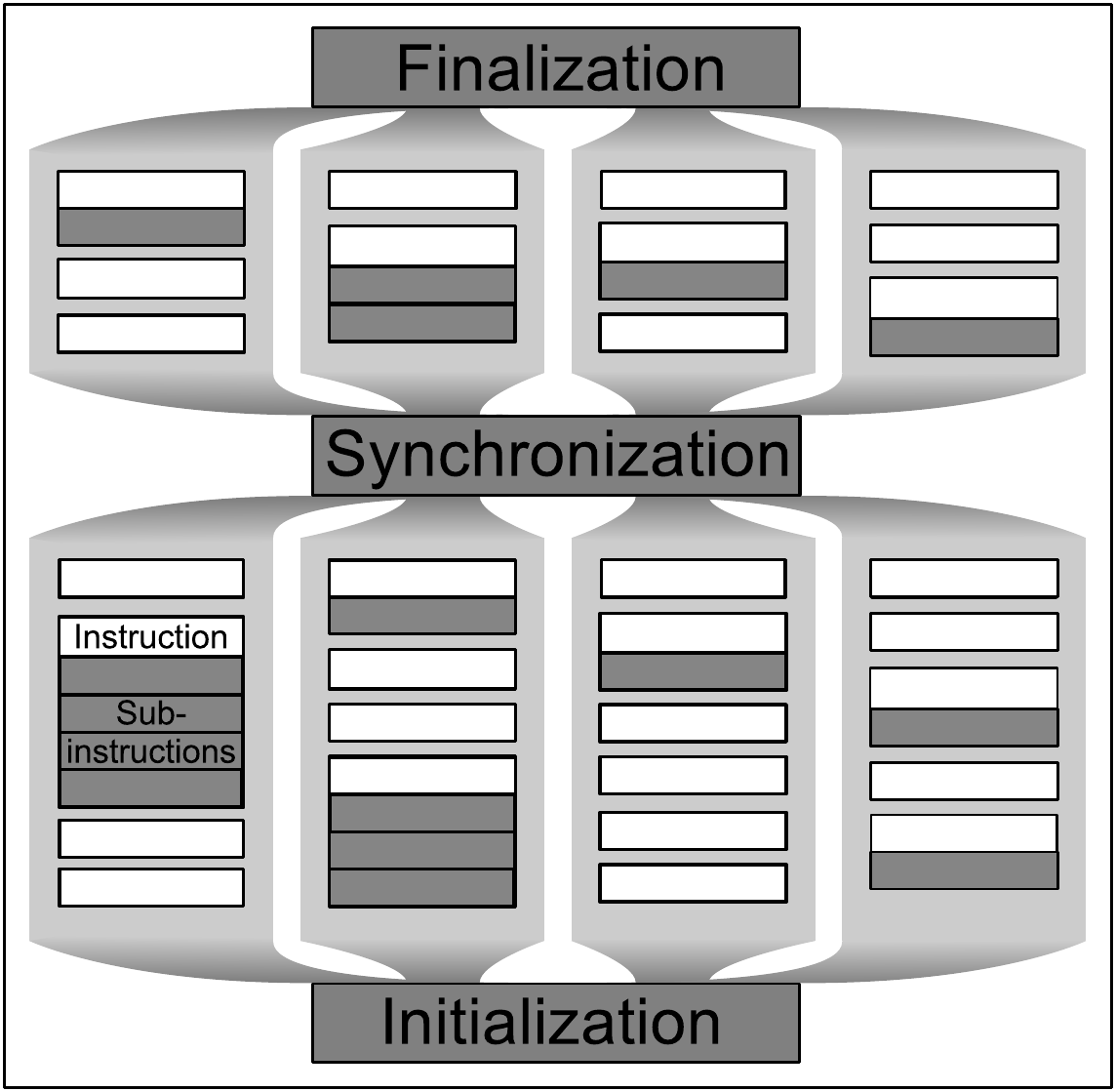}
\caption[Sketch of the parallelization scheme]{%
  Sketch of the parallelization scheme for \bytecode{} of two levels.
  Instructions and \subinstructions{} are in white and gray, respectively.
  Certain instructions imply that all following \subinstructions{} have to be
  executed before the next instruction is addressed. This grouping of
  instructions allows multiple sequential writes while minimizing
  synchronization points.
}
\label{fig:parallelizationscheme}
\end{figure}
%
The fact that we demand commutativity within a level implies that every
virtual register is changed by at most one thread.
A potential problem would hence be that the same address might be written to
successively multiple times in a computation though still being fully
disconnected to other parts.
To maintain the parallel nature \wrt{} the other parts and at the same time the
sequential nature of such a subcomputation, we can group instructions addressing
the same register to a building block.
A building block consists of one instruction and zero or more
\subinstructions{}.
\Subinstructions{} are conveniently represented in the \bytecode{} with negative
\acp{opcode} that are skipped over by the main loop.
Normal instructions can imply that all following \subinstructions{} have to be
executed sequentially before the thread computes the next instruction.
This is sketched in \cref{fig:parallelizationscheme}.
In \lang{OpenMP} this translates to a \code{parallel} region with a parallel
\code{do} loop over the instructions in a level.
More specifically, the complete parallelization of the \ac{VM} can be
written as
\begin{verbatim}
!$omp parallel
do level = 1, vm%N_levels - 1
   !$omp do schedule (static)
   do instruction = vm%levels (level) + 1, vm%levels (level + 1)
       call decode (vm, instruction)
   end do
   !$omp end do
end do
!$omp end parallel
\end{verbatim}
hereby we have an outer loop over levels as \code{vm\%levels} contains the
indices in the vector of instructions at which the level changes.
Due to our organization of the \bytecode{}, we can then perform the inner loop
in each level in parallel, whereby the \code{static schedule} just means that
every thread gets the same number of instructions.
The \code{decode} function is given the line number \code{instruction} in the
\code{instructions} block that should be translated.

As a side note, we want to mention that the sketched parallelization should be
very well suited for an implementation on a \ac{GPU}.
A common problem, encountered when trying to do scientific computing on a
\ac{GPU}, is the finite kernel size problem.
As noted \eg{} in \Rcite{MadGraphGPU2010}, large source code cannot be processed
by the \lang{CUDA} compiler, which is related to the fact that the numerous
cores on a \ac{GPU} are designed to execute simple operations very fast.
Dividing an amplitude into smaller pieces, which are computed one by one,
introduces more communication overhead
and is no ultimate solution since the compilation can still fail for complex
amplitudes~\cite{MadGraphGPU2010}.
The \ac{VM} on the other hand is a fixed small kernel, no matter how complex the
specific computation is.
A potential bottleneck might be the availability of the instruction block to all
threads, but this question has to be settled by an implementation and might have
a quite hardware dependent answer.

Finally, we note that the phase space parallelization mentioned in the beginning
of this subsection can still be applied.
When considering heterogeneous cluster or grid environments, where each node
is equipped with multi-core processors, a combination of distributed memory
parallelization for the combination of different phase space points and shared
memory parallelization of a single point seems to be a quite natural and
extremely potent combination.
\section{\OMEGA{} Virtual Machine}
\label{s:omega_virtual_machine}
The concept of a \ac{VM} can be easily applied to evaluate tree-level matrix
elements of arbitrary multiplicity.
The Optimizing Matrix Element Generator, \OMEGA{}~\cite{Omega2001}, avoids the
redundant representation of amplitudes in the form of Feynman-diagrams by using
\acp{1POW} recursively.
Just like the first two numerical codes \prog{ALPHA}~\cite{Caravaglios1995}
and \prog{HELAC}~\cite{HELAC2000}, which focussed on the \ac{SM}, \OMEGA{}
tames herewith the computational growth with the number of external particles
from a factorial to an exponential one but is completely general \wrt{} the used
Lagrangian.
Other programs with a very similar approach are \prog{Comix}~\cite{Comix2008},
based on the color-dressed Berends-Giele recursion formulation, and
\prog{Recola}~\cite{Recola2012}, which follows more closely the Dyson-Schwinger
formulation incorporating an important generalization~\cite{vanHameren2009} that
allows to compute one-loop amplitudes in the \ac{SM}.
\begin{table*}[htbp]
\caption{\Bytecode{} cheat sheet. Each instruction line consists of eight
  numbers having a different meaning depending on the first one, the
  \ac{opcode}. In general, the objects on the left hand side (lhs) are
  constructed from the right hand side (rhs). {X}, {Y} and {Z} are placeholders
  for the different Lorentz types of wave functions like fermions, scalars, etc.
  The value for width indicates which width scheme is used while its value and
the one of the mass is inferred from the \acs{PDG} code.\ outer\_ind denotes
spin and momentum index of the wave function.\ sym is the symmetry factor
computed from the number of identical particles involved.\vspace{1em} }
\label{tab:opcodes}
\centering
\begin{tabular}{l c c c c c c c}
  \toprule
    {{code}} & {coupl} & {coeff} &  {lhs} &
    $\T{rhs_1}$ & $\T{rhs_2}$ & $\T{rhs_3}$ & $\T{rhs_4}$ \\
  \midrule
    ADD\_MOMENTA & 0 & 0 & p\_lhs & p\_$\T{rhs_1}$ & p\_$\T{rhs_2}$ &
    p\_$\T{rhs_3}$ & 0 \\
    LOAD\_X & PDG & 0 & wf & outer\,\_\,ind & 0 & 0 & amp \\
    PROPAGATE\_Y & PDG & width & wf & p & 0 & 0 & amp  \\
    FUSE\_Z & coupl & coeff & lhs &
    $\T{rhs_1}$ & $\T{rhs_2}$ & $\T{rhs_3}$ & $\T{rhs_4}$ \\
    {CALC\_BRAKET} & sign & 0 & amp & sym & 0 & 0 & 0 \\
  \bottomrule
\end{tabular}
\end{table*}
%

The model-independence is achieved in \OMEGA{} with the meta-programming ansatz
mentioned earlier whereby the symbolic representation is determined in
\lang{OCaml}.
This abstract expression is then translated to valid \lang{Fortran} code that
is automatically compiled and used in \prog{Whizard}~\cite{Whizard2011} for
event generation.
As \OMEGA{} has been designed in a modular way, it has been rather
straightforward to add an additional output module that produces \bytecode{}
instead of \lang{Fortran} code.
Some additional technical details about the implementation can be found
in \Rcite{bcn_master} and more completely in the documented source
code~\cite{omega_trunk}.

The number of distinct operations that have to be performed in the computation
of a cross section is related to the Feynman rules and therefore quite limited.
As such, these operations are very good candidates for the translation to
\bytecode{}.
In fact, this results in only about 80 different \acp{opcode} for the
complete \ac{SM}, which have been implemented in the \ac{OVM}.
In order to support completely general Lagrangians with arbitrary
tensor structures as in \cite{UFO2012,Aloha2012},
the subroutines implementing the vertices can be
mapped to \acp{opcode} dynamically.
They can be classified as described in \cref{tab:opcodes} as ADD\_MOMENTA,
LOAD\_X, PROPAGATE\_Y, FUSE\_Z and CALC\_BRAKET, \ie{} the addition of momenta,
the construction of external wave functions, the propagation of wave functions,
the fusion of wave functions according to the Feynman rules and the computation
of the final braket, which yields the amplitude with appropriate prefactors,
respectively.
This limited set of instructions as well as the objects in a calculation can
each be identified unambiguously with an integer.
To obtain this integer in \OMEGA{}, we apply a map from a given set of objects,
\eg{} wave functions, to the numbers from 1 to $N$, where $N$ is the cardinality
of the set, by using an ordering that ensures that distinct objects are
not assigned the same number.
To discriminate between particle flavors, there is of course already a well-known
ordering that we can use, namely the \ac{PDG} integers~\cite{PDG2014}.

\begin{figure}[htbp]
\centering
\includegraphics[width=.7\columnwidth,height=\picheight,keepaspectratio]
{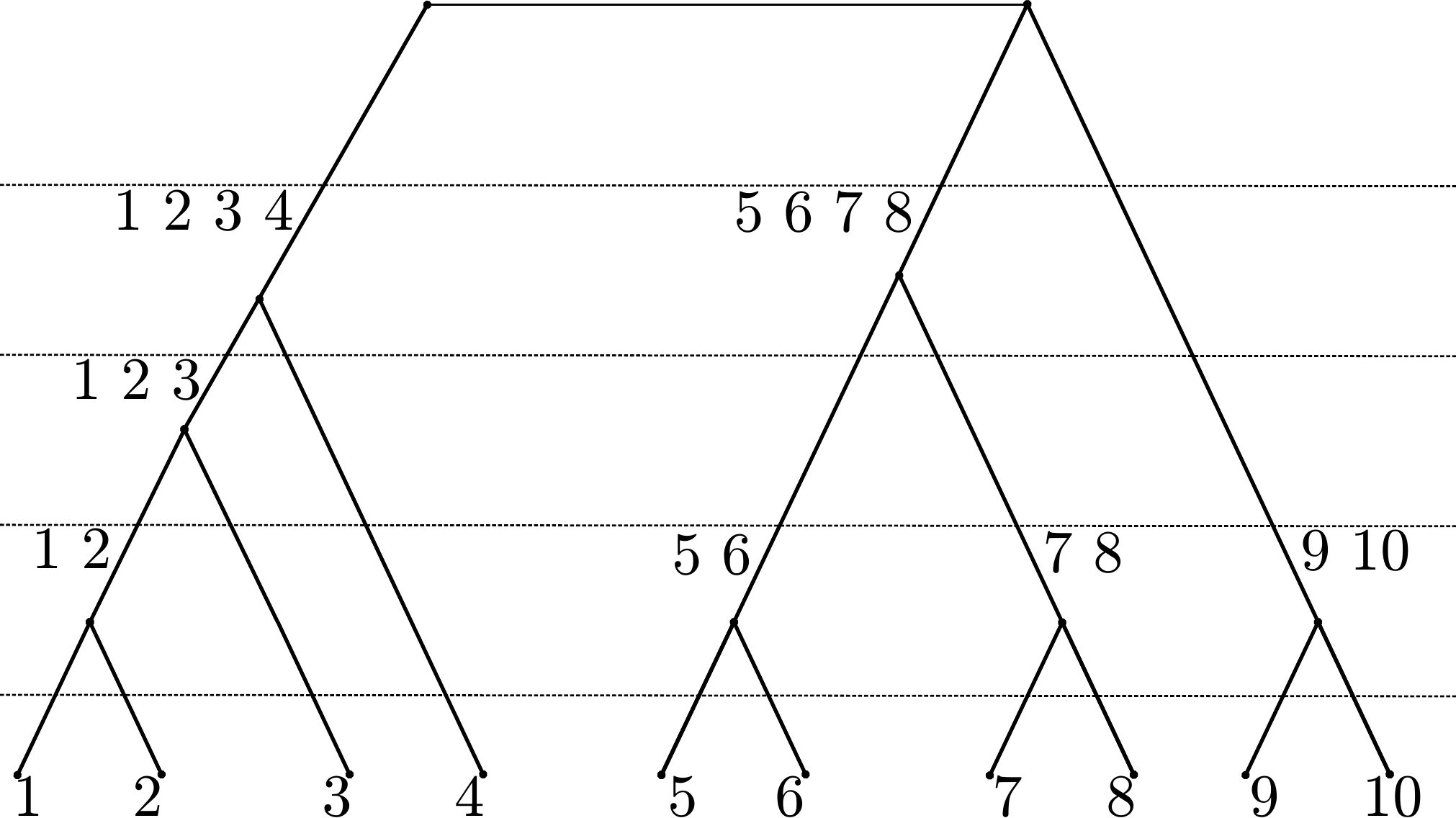}
\caption[Classification of levels by the number of summands in the
momenta]{The classification of levels by the number of summands in the momenta
yields an unambiguous organization of the calculation whereby each level can
be calculated in parallel. We emphasize that this illustration is only one of
thousands of possible partitions, whereby each \ac{1POW} is heavily reused.}
\label{fig:momentalevels}
\end{figure}

For the parallel execution, we identify the different levels by the number of
external momenta a wave function is connected to or equivalently the number of
summands in the momentum of the wave function.
This is depicted in \cref{fig:momentalevels}.
Furthermore, we have to group the FUSE\_Z instructions to building blocks
together with either PROPAGATE\_Y or CALC\_BRAKET instructions.
In this sense, all FUSE\_Z instructions are \subinstructions{} that can belong
to either of these two building blocks and the \ac{OVM} will either form a
\ac{1POW} $\phi(p+q)=\phi(p)\phi(q)$ or the amplitude
$\mathcal{A}=\phi(p)\phi(q)$ depending on the main instruction.
%

The \ac{OVM} is initialized with a call that specifies where to find the
\bytecode{} file, what versions of the \ac{OVM} and physics model are used, as
well as input arrays for masses, widths and couplings, which hold the numeric
values for the different types of particles and interactions.
In the header of the \bytecode{} file the \ac{OVM} finds the number of momenta,
amplitudes (due to multiple color flows and flavor combinations) and wave
functions that should be allocated.
This is followed by fixed tables for spin, flavor, color flows and color ghosts,
for details concerning the color flow formulation \cf{} \Rcite{KORS2012},
as well as whether a certain flavor-color combination is allowed.
Finally, the body of instructions completes the necessary information to compute
the cross section.
\section{Speed Benchmarks}
\label{s:benchmarks}
In this section, we benchmark the \ac{OVM} against the compiled code in
\cref{ss:runtime_performance}, analyze
the scaling behavior with multiple cores, which indicates to which degree we
are computing in parallel and how much speed up we can expect for more cores, in
\cref{ss:parallelization} and end with a remark on the \bytecode{} generation
performance in \cref{ss:bytecode_generation}.
All processes shown here, and various others, have been validated against the
compiled versions where possible for random massless momenta, generated by
\prog{Rambo}~\cite{Rambo1986}, with the help of an automated test suite that is
run when \code{make}~\code{check} is started in the build folder of \OMEGA{}.
Further tests or benchmarks can be added by appending a single line to
the two steering files.
We stress that every process is computed in its respective model (QED, QCD or
SM) to \emph{full} tree-level order including all interferences and we have not
restricted \eg{} the Drell-Yan amplitudes to only one electroweak propagator.
For simplicity of the test and benchmark suite, we use massless momenta
but are in no way restricted to massless theories and do not use simplifications
that would render the massless code faster.

To avoid complete compiler dependence of the results, we use two different
compilers that are commonly used in scientific projects.
These are the GNU and Intel compilers, \code{gfortran 4.7.1} and \code{ifort
14.0.3}, respectively.
We do not claim that our results are necessarily representative for all
\lang{Fortran} compilers or even compiler versions, but they should still give a
good impression of the expected variance in performance.
For multi-threading, we use the \prog{OpenMP} library of the compilers as we
are only interested in shared memory parallelization as discussed in
\cref{ss:general_parallelization}.
The evaluation time measurements are performed on a computer with two {Intel(R)
Xeon(R) E5-2440 @ 2.40GHz} \acp{CPU}, having \SI{16}{\mibi\byte} L3 cache on each
socket, and 2x \SI{32}{\gibi\byte} RAM running under {Scientific Linux 6.5}.
The machine has been locked down exclusively for these runs to minimize context
switches as far as possible.
\subsection{Runtime Performance}
\label{ss:runtime_performance}
\begin{figure}[htb]
\centering
\includegraphics[width=\columnwidth,height=\picheight,keepaspectratio]
{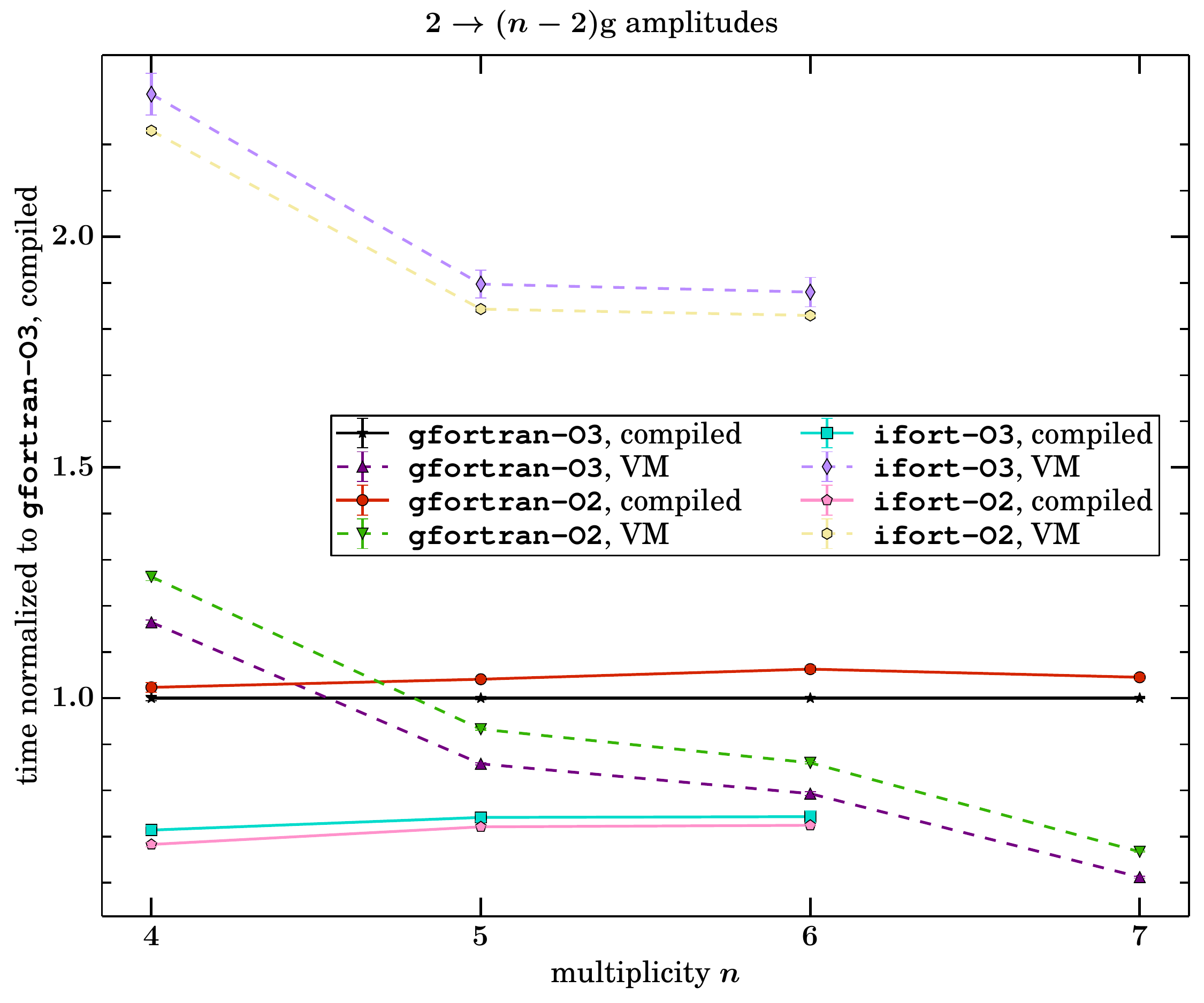}
\caption{\ac{CPU} times measured with the \lang{Fortran} intrinsic
  \code{cpu\_time} and normalized for each process to the compiled source code
  using \code{gfortran -O3}.  Dashed (solid) lines represent the \ac{OVM}
  (compiled source code). The error bars correspond to the standard deviation of
  three runs.}
\label{fig:g.pdf}
\end{figure}

\begin{figure}[p]
\centering
\includegraphics[width=\columnwidth,height=\picheight,keepaspectratio]
{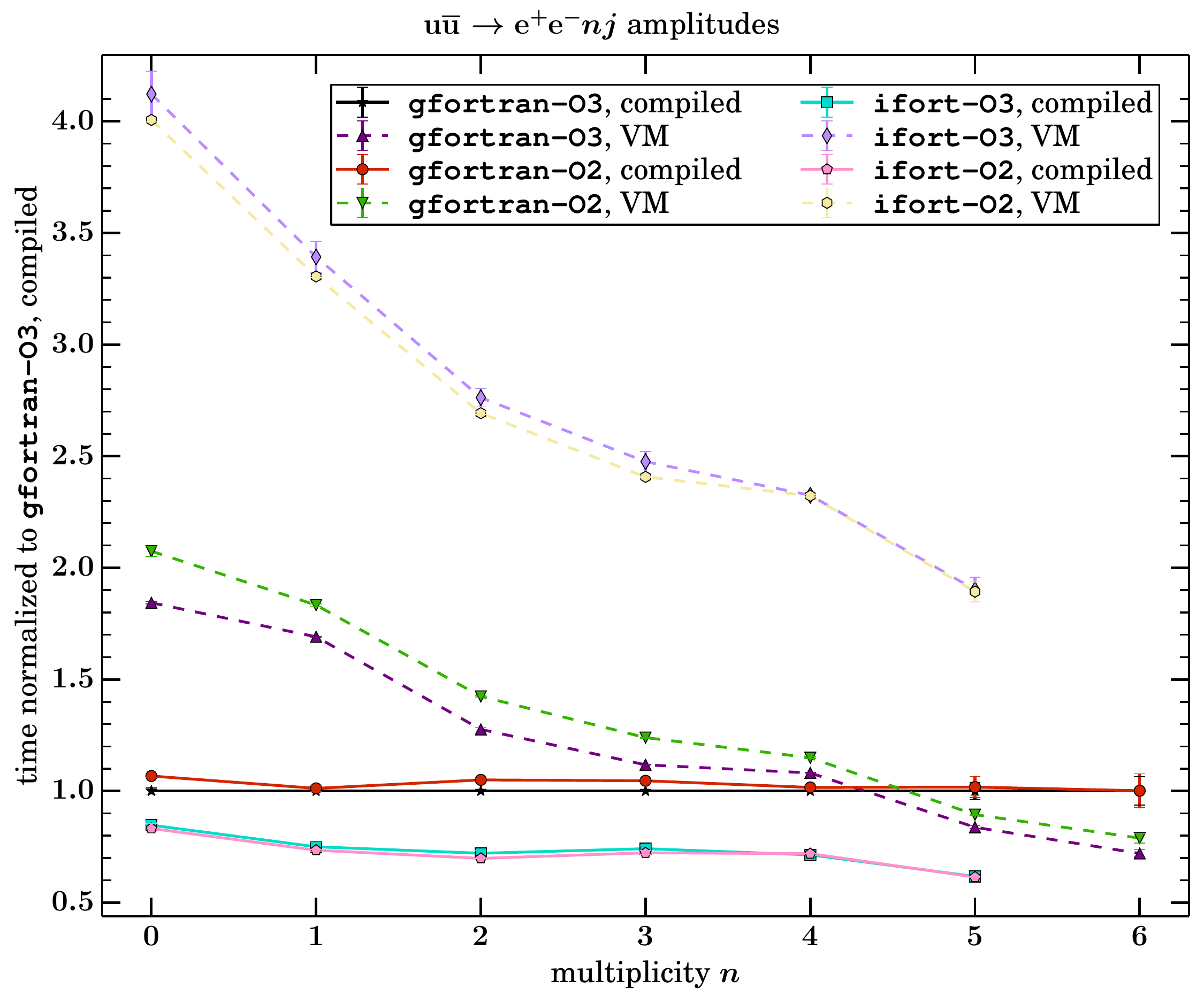}
\caption{Same as \cref{fig:g.pdf} but for the \ac{SM} Drell-Yan process $\Pqu
\Paqu \to \Ppositron \Pelectron n j$ where $j=\Pqu,\Paqu,g$.}
\label{fig:DY.pdf}
\end{figure}

\begin{figure}[p]
\centering
\includegraphics[width=\columnwidth,height=\picheight,keepaspectratio]
{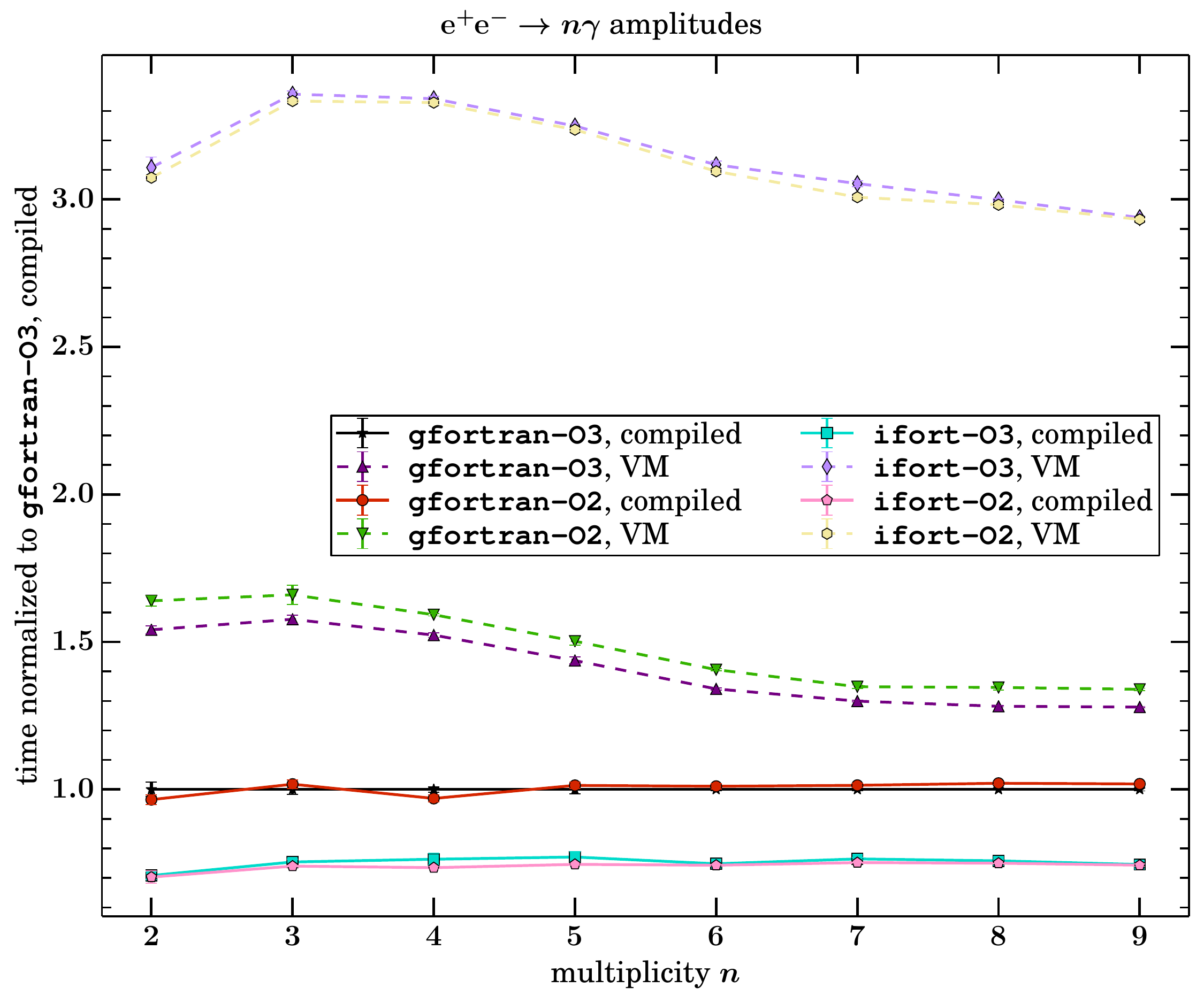}
\caption{Same as \cref{fig:g.pdf} but for \ac{QED} photon production $\Ppositron
\Pelectron \to n\Pphoton$.}
\label{fig:ee.pdf}
\end{figure}

In Figure~\ref{fig:g.pdf},~\ref{fig:DY.pdf} and~\ref{fig:ee.pdf}, we show the
measured \ac{CPU} times for QCD, SM and QED processes with two different
optimization levels for the compiled code and the \ac{OVM} using the GNU and
Intel compiler.
Since the evaluation times are highly reproducible, we use only three runs to
obtain mean and standard deviation.
In most cases this results in vanishing error bars.
We stress that we show here the relative times normalized for each process to
\cd{gfortran-O3}, which is why the times are not growing with the number of
particles.
Absolute times for fully color and helicity summed amplitudes are increasing at
least like $2^n$ due to helicity and like $(n-1)!{}$ (for the gluon amplitude)
due to the number of color flows if no Monte Carlo methods are employed to
include these sums in the integration.
Lower optimization levels than \cd{-O2} are not competitive in terms of run
time.
For \cd{gfortran}, we observe for most processes the fastest performance
with \cd{-O3} and for \cd{ifort} with \cd{-O2}, which is an effect commonly
encountered.
The fastest performance is given by the source code compiled with \cd{ifort-O2}
being roughly \num{0.75} times the time needed by \cd{gfortran-O3}.
%

The crucial point, however, is that \cd{ifort} fails to compile the $n=7$ gluon
and the $\Pqu\Paqu\to\Ppositron\Pelectron6j$ Drell-Yan process while the \ac{OVM}
immediately starts computing.
The GNU compiler is usually able to compile one multiplicity higher compared
to the Intel before breaking down.
This fits together with the better performance of the compilable processes and
longer compile times as \cd{ifort} seems to apply more sophisticated
optimization methods to the source code.
Disabling the optimizations with \cd{-O0} still does not allow to compute the
aforementioned processes with both compilers.

Another interesting observation is that the \ac{OVM} gets faster compared to the
compiled code with increasing multiplicity of external particles though this
feature is more pronounced in SM and QCD processes.
This is no initialization effect since we allocate the arrays in the beginning
and only measure the generation time of matrix elements for $M$ different phase
space points.
$M$ has been set beforehand for each process with the known approximate scaling
for higher multiplicities such that it takes a couple of minutes to complete the
computation to have a reliable measurement.
The absolute costs for translating an instruction line to actual machine code,
\ie{} the virtualization costs, are proportional to the number
of instructions resulting hence in a constant factor in the relative,
normalized time and can not account for this scaling behavior.
The most important difference between the compiled source code and the \ac{VM}
is then the explicit double loop in the \ac{VM}, which goes over the
instructions in a level and over all levels as shown in the code excerpt in
\cref{ss:general_parallelization}.
The advantages and disadvantages of the double loop are basically the same as
general loop unrolling considerations.
The native source code represents hereby the unrolled loop that does not have to
check for the loop variables, can use latency hiding to start the next
instruction while waiting for memory, potentially use \ac{CSE}\footnote{Although
  all common subexpressions have in our case already been avoided by \OMEGA{}.}
and optimize the prefetching of the processor.
The double loop of the \ac{VM} on the other hand has the advantage of having a
higher probability to keep the decode function in the instruction cache.
This can potentially explain the scaling behavior with growing complexity
compared to the compiled code.
We observe roughly the same effect for both compilers, but the \ac{OVM}
compiled with \cd{ifort} is about a factor of two slower than the version
with \cd{gfortran}, rendering it not really useful for production runs.
This could eventually be solved with a profile-guided optimization, but this is
beyond the scope of this work.
%

Finally, we want to understand the performance difference of the \ac{OVM}
between the QED and QCD amplitudes.
To get an impression of the computational complexity, consider that the
$\Ppositron\Pelectron\to9\gamma$ amplitude is represented by
\SI{125}{\kibi\byte} and the $\Pgluon\Pgluon\to4\Pgluon$ by
\SI{269}{\kibi\byte} of \bytecode{} consisting of 3373 and 6780 instructions,
respectively.
Since the complexity grows exponentially with the number of external particles,
those processes can be considered approximatively equally expensive.
The difference is, however, that the QED amplitude consists, due to the very
high number of external particles, of 8 levels while the QCD amplitude has only
4 levels.
This results in about 422 and 1695 instructions per level on average, which is
why we can expect worse parallel performance for the QED amplitude due to
higher synchronization costs compared to the work to be done per level.
Furthermore, this is accompanied with higher memory needs:
For the QED amplitude, we need 549 momenta, 256 spinors, 256 conjugated spinors
and 9 vector wave functions.
The QCD amplitude, on the other hand, requires only 31 momenta and 330 vector
wave functions.
Returning with this information to the argument made in the last paragraph, the
compiled code can gain more from data prefetching in the case of the QED
amplitude while the \ac{VM} improves for more instructions on less data as it is
the case for QCD due to the lower chance for instruction cache misses.
Overall, we can expect QED to be the worst case scenario for the \ac{OVM} as it
has the lowest number of flavors and the simplest gauge structure one can think
of\footnote{Excluding toy models like $\phi^4$ theory.}.
Considering all results, we find that the runtimes are in the same order of
magnitude and that a \ac{VM} can be competitive in terms of speed with the
compiled version, especially for extreme computations with a high amount of
operations per memory.
\subsection{Parallelization}
\label{ss:parallelization}
%
Amdahl's idealized law~\cite{amdahl} simply divides an algorithm into
parallelizable parts $p$ and strictly serial parts $1-p$. Therefore, the
possible speedup $s$ for a computation with $n$ processors is
\begin{align}
  s(n) \equiv \frac{t(1)}{t(n)} = \frac 1{(1-p) + \frac p n} \po
\label{eq:amdahl}
\end{align}
\begin{figure}[htb]
\centering
\includegraphics[width=\columnwidth,height=\picheight,keepaspectratio]
{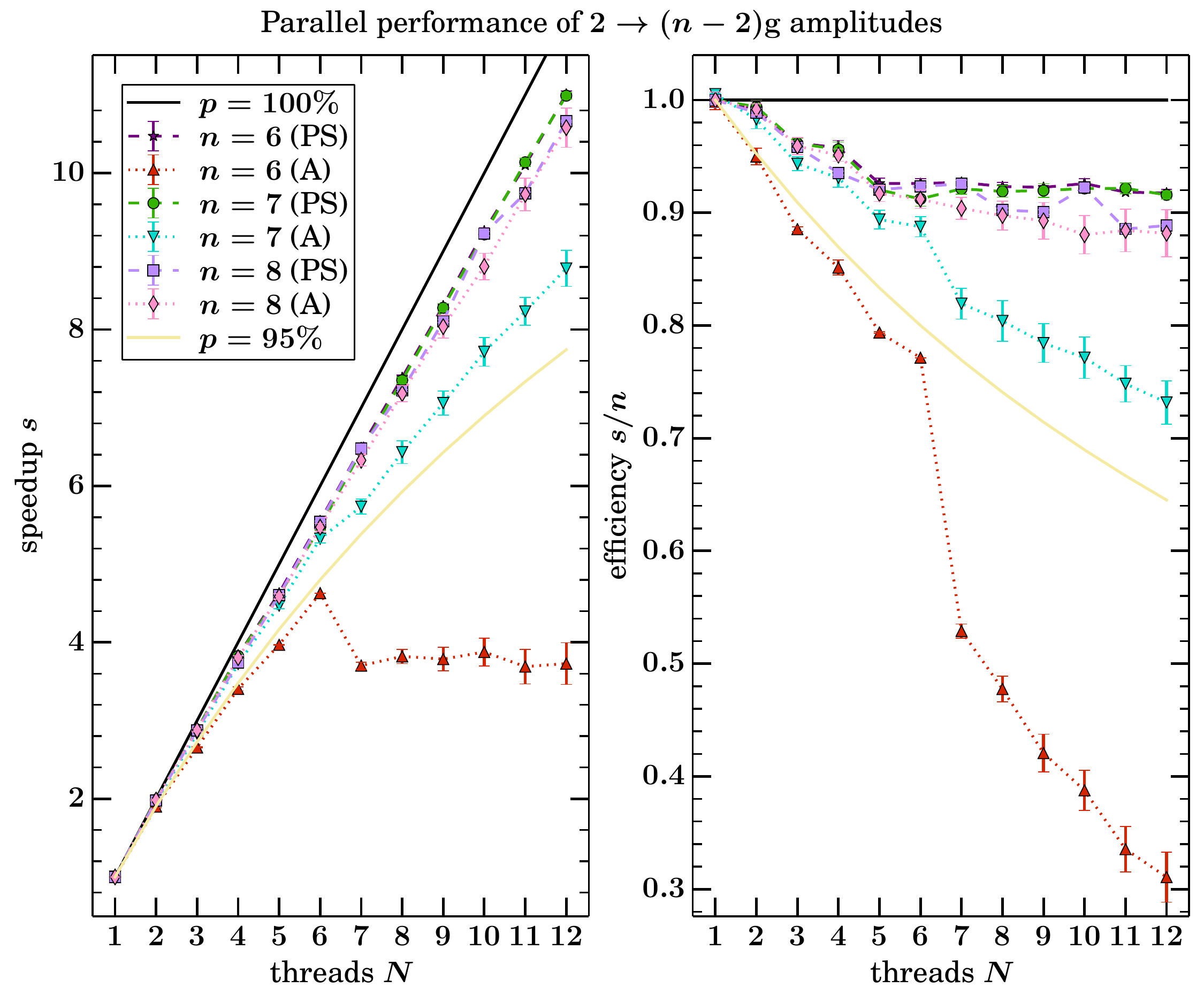}
\caption{Speedup and efficiency to compute a fixed number of phase space points
  for the parallel evaluation of multiple phase space points (PS) and the
  parallel evaluation of the amplitude itself (A) are shown as dashed and dotted
  lines. The error bars correspond to the standard deviation of three runs. The
  solid lines represent Amdahl's law for a fixed value of the parallelizable
  part $p$.}
\label{fig:par_g.pdf}
\end{figure}
Communication costs between processors $\order{n}$ have been neglected hereby in
the denominator of \cref{eq:amdahl}.
This means that we have $\lim_{n\to\infty}s(n)=1/(1-p)$ in the idealized case and
$\lim_{n\to\infty}s(n)=0$ including communication costs.
In reality, we are interested in high speedups for finite $n$ and also have to
care about efficient cache usage.
The picture becomes more complicated in modern \ac{NUMA} environments with
multiple \acp{CPU} on the same board where each socket has its own memory
that the others can access as distributed shared memory.
For our machine, the two sockets have even and odd numbers for the \acp{CPU} on
them.
To improve the thread scheduling, we have pinned the threads to the cores via
\begin{verbatim}
GOMP_CPU_AFFINITY='0 2 4 6 8 10 1 3 5 7 9 11'
\end{verbatim}
corresponding to using the first \ac{CPU} for the threads 1--6 and then the
second for 7--12.
Hyper-threading is disabled as it is not expected to speedup such a calculation.
Sadly, we could not achieve any $s>1$ for the parallelization of the \ac{OVM}
with the Intel compiler neither by using multiple phase space points at once nor
by computing the amplitude in parallel.
The reason for this is quite unclear, as the exact same code shows the expected
speedup with the GNU compiler, and seems to be correlated with the bad
single-core performance of the \ac{OVM} compiled with \code{ifort}.

\begin{figure}[p]
\centering
\includegraphics[width=\columnwidth,height=\picheight,keepaspectratio]
{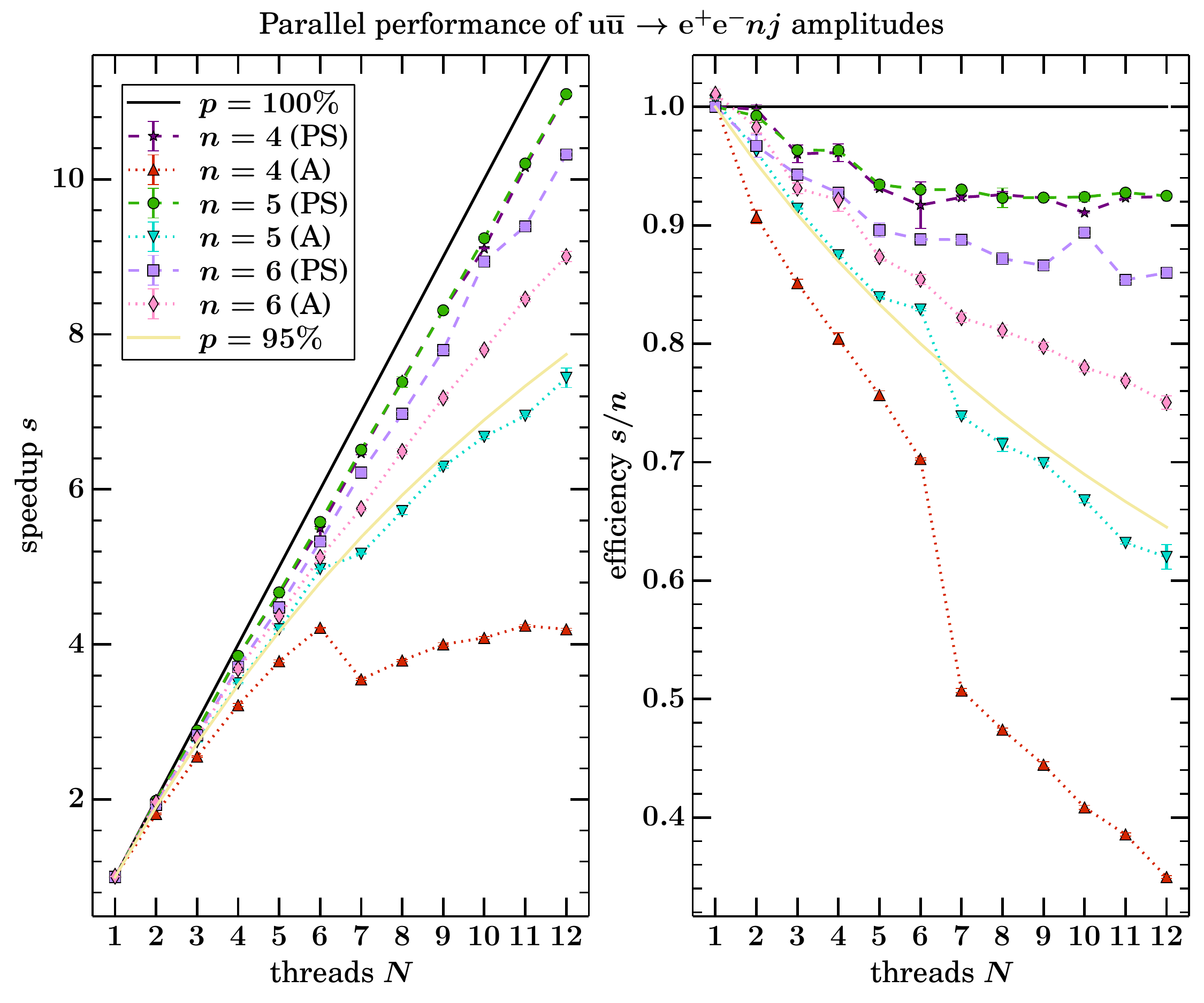}
\caption{Same as \cref{fig:g.pdf} but for the \ac{SM} Drell-Yan process $\Pqu
\Paqu \to \Ppositron \Pelectron n j$ where $j=\Pqu,\Paqu,g$.}
\label{fig:par_DY.pdf}
\end{figure}
\begin{figure}[p]
\centering
\includegraphics[width=\columnwidth,height=\picheight,keepaspectratio]
{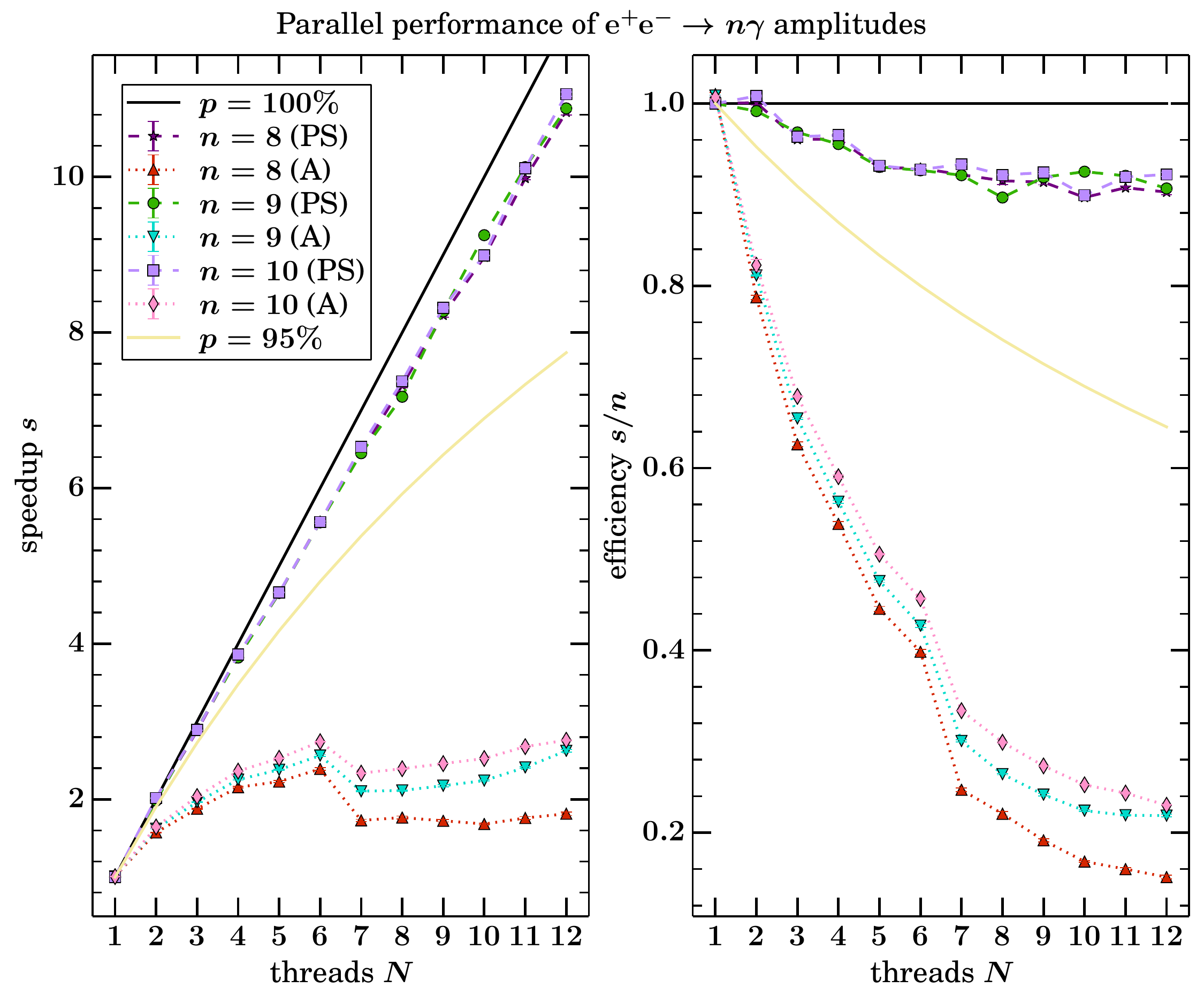}
\caption{Same as \cref{fig:g.pdf} but for \ac{QED} photon production $\Ppositron \Pelectron \to n\Pphoton$.}
\label{fig:par_ee.pdf}
\end{figure}
In Figure~\ref{fig:par_g.pdf},~\ref{fig:par_DY.pdf} and~\ref{fig:par_ee.pdf}, we
show the speedup with multiple cores $N$ by either using the parallelization
procedure, discussed in \cref{ss:general_parallelization}, to compute one
amplitude in parallel or by computing multiple amplitudes for multiple phase
space points in parallel again for processes with different multiplicities $n$
in QCD, SM and QED\@.
In a real application the phase space parallelization can not be as efficient as
the naive version here, where we can just parallelize the \code{do} loop over
$N_\T{points}$, since usually \prog{Vegas}~\cite{Vegas1976} grids are used to
approximate the matrix-element and these have to be adjusted iteratively.
These book-keeping tasks reduce the parallel parts and the phase-space
parallelization shown here (PS) can therefore be regarded as upper bounds.
For the parallelization, we chose to only compare the runtime of a single
helicity combination to reduce the overall time needed to perform the tests
since numerical off-shell recursion algorithms have the same runtime for every
helicity, opposed to the closed analytical formulae~\cite{Badger2013}.
To measure the speedup we have used wall clock times as given by the OpenMP
function \path{omp_get_wtime}.
In \cref{fig:par_g.pdf}, we can see that the $n=7$ and $n=8$ gluon amplitudes
parallelize very well with both methods with parallelizable parts above
$\SI{95}{\percent}$.
In the shared memory parallel evaluation of the amplitude (A), the impact of the
architecture is quite obvious.
For $N=7$, \ie{} when the second \ac{CPU} of the \ac{NUMA} environment is
activated, we see a sharp drop in efficiency, which can be expected since there
will be synchronization costs at the end of each level and costs to maintain
cache coherency after each instruction inside the amplitude.
For the $n=6$ gluon and the $n=4$ Drell-Yan process, this effect even leads to a
saturation in the speedup indicating that the performance is bound by the memory
transfer rates between both \acp{CPU}.
We observe that this effect becomes less important for more complex amplitudes.
To understand this, note that Sandy Bridge with its Intel Quick Path
Interconnect (QPI) is actually a cache coherent \ac{NUMA} architecture, meaning
that the cache controllers are required to maintain a consistent memory image
when more than one local cache stores the same memory location.
Such cache coherency effects have been studied \eg{} for the related Nehalem
microarchitecture in \Rcite{Molka09}.
They have shown that the bandwidth to other cores strongly depend on the
coherency state of the accessed data.
If the latest copy is in the local caches of the remote core, which is more
likely to occur for smaller processes, read bandwidths decrease significantly.
As expected by the discussion in \cref{ss:runtime_performance}, the QED
amplitudes do only parallelize well if phase space parallelization is used.

The very good performance of the phase space parallelization can be explained by
the available cache.
The size of the L3 cache per core, \SI{2.7}{\mibi\byte}, is more than enough to
host $N=12$ independent versions of the \ac{OVM} even for the $n=8$ gluon
amplitude, where momenta, amplitudes and wave functions account to
\SI{464.77}{\kibi\byte}.
This will break down for this architecture, however, for one multiplicity higher
if we extrapolate the given scaling for the number of objects involved in the
calculation.
The current version of \OMEGA{} will produce code for \emph{all} color
flows of a given process simultaneusly.  Therefore we have not
included the $n=9$ gluon amplitude in the tests, because the
$(9-1)!=40\,320$ different color flow amplitudes do not fit into
memory.  For real world applications the summation of all color
amplitudes will have to be replaced by a sampling of color space.

Either way, it is important to also have the possibility to compute one
amplitude in parallel since architectures change and \eg{} the Intel Xeon Phi
has only $\SI{512}{\kibi\byte}$ cache per core, rendering already the $n=8$ case
close to inappropriate for phase space parallelization.
\subsection{Bytecode Generation}
\label{ss:bytecode_generation}
It is intuitively clear that integer \bytecode{} is smaller than syntactically
correct \lang{Fortran} source code.
Furthermore, we use long strings in the source code for debugging purposes,
\ie{} to directly see to which color flow and momentum combination a \ac{1POW}
belongs.
To be specific, we note that the \bytecode{} for the \ac{OVM} is about one order
of magnitude smaller.
For convenience, some values together with their old compile times are shown in
\cref{tab:ovm_bc_creation}.
The bytecode size has been furthermore almost halved for very colorful
amplitudes in a later version, by using the symmetry of the color factor
table, but this could have been achieved with the \lang{Fortran} output as well
and is not shown here.
The smaller output format leads to less required RAM and time to produce it.
Especially for many color flows, where the generation time of \OMEGA{} is
dominated by the output procedure, we observe \eg{} for
$\Pgluon\Pgluon\to6\Pgluon$ a reduction in memory from
\SI{2.17}{\gibi\byte} to \SI{1.34}{\gibi\byte} and in generation time from
\SI{11}{\minute}~\SI{52}{\second} to \SI{3}{\minute}~\SI{35}{\second}, while
staying roughly the same for small processes.
\begin{table}[htbp]
  \caption{Size of the \bytecode{} (\cd{BC}) compared to the \cd{Fortran} source
    code together with the corresponding compile time with \cd{gfortran}. The
    compile times were measured on a computer with an \hardware{i7--2720QM}
    \ac{CPU}.
    The $2\Pgluon\to6\Pgluon$ process fails to compile.
    \vspace{1em} }
\label{tab:ovm_bc_creation}
  \centering
  \begin{tabular}{l c c c}
    \toprule
    process & \cd{BC} size & \cd{Fortran} size & $t_\T{compile}$ \\
    \midrule
    $\Pgluon \Pgluon \to \Pgluon \Pgluon \Pgluon \Pgluon \Pgluon \Pgluon$ &
    \SI{428}{\mibi\byte} & \SI{4.0}{\gibi\byte}
    & - \\
    $\Pgluon \Pgluon \to \Pgluon \Pgluon \Pgluon \Pgluon \Pgluon$ &
    \SI{9.4}{\mibi\byte} & \SI{85}{\mibi\byte}
    & \SI{483+-18}{\second} \\
    $\Pgluon \Pgluon \to q \bar{q} q'\bar{q}' q''\bar{q}'' g$ & \SI{3.2}{\mibi\byte} &
    \SI{27}{\mibi\byte} & \SI{166+-15}{\second} \\
    $e^+ e^- \to5\,(e^+ e^-)$ &
    \SI{0.7}{\mibi\byte} & \SI{1.9}{\mibi\byte} & \SI{32.46+-0.13}{\second} \\
    \bottomrule
  \end{tabular}
\end{table}
\section{Summary and Outlook}
\label{s:conclusions}
%
A \ac{VM} circumvents the compile and link problems that are associated with
huge source code as it emerges from very complex algebraic expressions.
This work is a, to our knowledge first, proof of principle that \acp{VM} are
indeed a viable option that is maintaining relatively \highperformance{} in the
numerical evaluation of these expressions and allows to approach the hardware
limits.
In practice, a \ac{VM} saves hours of compile time that would result often enough
in internal compiler errors instead of working code.
The concept has been successively applied to construct the \ac{OVM} that is now
an alternative method to compute tree-level matrix elements in the publicly
available package \OMEGA{} and will also be integrated in \prog{Whizard} in an
upcoming release of the package.
Any computation can in principle be performed with a \ac{VM} though the benefits
are clearly in the regime of extreme computations that are not solvable with the
conventional method.
Here, we have seen that \acp{VM} can even perform better than compiled code.
Also the parallelization of the amplitude is for very complex processes close to
the optimum.

It would be an interesting experiment to reduce the virtualization overhead by
using an actual machine to compute matrix elements.
The number of instructions corresponding to different wave function
fusions and propagators is finite for renormalizable theories
(including effective theories up to a fixed mass dimension) and
implemented similarly in the various matrix element generators.  If the
authors can agree on a common set of instructions and conventions this
machine could therefore be used by all those programs.
The \ac{LHC} collaborations might actually have a need for this, especially in
the light of the \ac{HL-LHC}, where the number of events for simulation and
reconstruction increases by an order of magnitude and new computing clusters
will most likely be needed.
\acp{FPGA} can serve as such a machine as they have comparable if not superior
floating-point performance with respect to current microprocessors and the
\ac{OVM} and its instruction set is the first step to test the feasibility and
potential gains of computing matrix elements in this environment.
The hardware integration might be quite easy as Intel has recently
revealed~\cite{IntelFPGA2014} that Xeon processors can in future be paired
with a \ac{FPGA} in a single socket.

While \acp{GPU} and \acp{FPGA} are rather unconventional devices that will
need large code modifications, similar speedups could be achieved with
the \ac{MIC} platform.
Various existing scientific applications in \lang{Fortran} and \lang{C++} have
been analyzed in \Rcite{TACCMIC2012} with an early development environment
release of the upcoming Intel Xeon Phi.
They have shown that it is possible to compile libraries that utilize the
\lang{Autotools} build system for the \ac{MIC} environment just by setting the
proper \code{./configure} options, at least for static builds.
This is a clear advantage as no rewriting is necessary while the speedup can
still be in the order of $20$ for about 100 threads.
Obviously, this strongly depends on having a highly parallel code.
We would expect for the \ac{OVM} speedups in the range of $17-50$ for processes
that exhibit \SIrange{95}{99}{\percent} parallel fractions by extrapolating the
data of \Cref{s:benchmarks} and assuming no severe memory problems.
In fact, the Xeon Phi possesses no L3 cache at all but a set of coherent
L2 caches with less overall cache per core.
Thus, we might see a break down in the efficiency of the phase space
parallelization, when the objects of one matrix element exceed the L2 cache,
while on the other hand high speedups in the parallelization of the amplitude
can be maintained.

\subsection*{Acknowledgments}
BCN thanks Danny van Dyk, Tomas Jezo and Jos Vermaseren for useful discussions
of the idea.
\appendix
\section{A Trivial Example}
\label{a:a_trivial_example}
To try out the ideas of \cref{s:virtual_machine}, we evaluate a trivial problem
here.
Consider the identity, for $x\in\R$,
\begin{align}
  (-1)^x &= \E^{\I\,\pi\,x} + \E^{\ln{2}} + \frac 1{1-\frac 12} - \frac 1{(1-\frac
  12)^2} \no
  &\equiv C_1(x) + \E^{R_1} + R_2 - R_3 \co{}
  \label{eq:toy}
\end{align}
which can be written in terms of the known series $C_1$ and $R_i$ as
\begin{align}
  C_1(x) &= \sum_{n=0}^\infty \frac 1{n!} \left(\I\,\pi\,x\right)^n &
  R_1 &= \sum_{n=1}^\infty (-1)^{n+1} \frac 1 n \no
  R_2 &= \sum_{n=0}^\infty \left(\frac 1 2\right)^n &
  R_3 &= \sum_{n=1}^\infty n \left(\frac 1 2\right)^{n-1} \po
  \label{eq:all_terms}
\end{align}
To test the \ac{VM}, we can compute these series explicitly
numerically.
This means we create \bytecode{} for a given $N$, whereby the
above equations follow for $N\to\infty$, and execute it in the \ac{VM}.
Though these series are not particularly interesting by themselves, it allows
us to test the whole \ac{VM} infrastructure in a self-contained way, \ie{} without
dependencies on external libraries\footnote{Except \lang{OpenMP}, which is needed
for parallelization, but the single-threaded execution also works without the
library.}.
The corresponding template code is freely accessible at
\url{https://github.com/bijancn/basic-vm} and can be used to create a
\ac{VM} for any purpose.
It is written in a subset of \lang{Fortran2003} that is supported by most modern
compilers for mere convenience of the author and due to the environment in which
the \acf{OVM} is used.
A translation to \lang{C} or \lang{Fortran95} is straightforward as the code
structure is very simple.
An earlier version in \lang{Fortran95} had the same performance characteristics
in the tested cases as the one shown here, indicating that possible performance
penalties for the use of some higher-level constructs on the top-level are
negligible.

The first step in writing a \ac{VM} is to identify the set of operations that
are needed to perform the computation.
Let us associate an instruction line of five integers as \acf{opcode}, \ac{LHS}
and \ac{RHS}
\begin{verbatim}
  OPCODE LHS RHS1 RHS2 RHS3.
\end{verbatim}
The three fundamental operations, identified by the \acp{opcode} 1-3, could then
be
\begin{description}
  \item[1] \verb?tmp_real(LHS) += RHS1 / RHS2?
  \item[2] \verb?tmp_cmplx(LHS) += (const(RHS1) * input(RHS1))^RHS2 / table(RHS3)?
  \item[3] \verb?output(LHS) = tmp_cmplx(LHS) + exp(tmp_real(RHS1))? \\
    \verb? + tmp_real(RHS2) - tmp_real (RHS3)?.

\end{description}
Hereby, we have chosen to put the factorials needed into a separate
\code{table}. These are given in the \bytecode{} as a simple line-by-line array
\begin{verbatim}
  1
  1
  2
  6...
\end{verbatim}
The calling application has to supply the constant block
\begin{align}
  \mathtt{const(1)}\;&=\;\I\,\pi \intertext{and as input data}
  \mathtt{input(1)}\;&=\;x.
\end{align}
The \bytecode{} for the first elements of $C_1=1+\I\pi-\frac 12\pi^2+\dots$ now
reads, \eg{},
\begin{verbatim}
  2 1 1 0 1
  2 1 1 1 2
  2 1 1 2 3
  2 1 1 3 4
\end{verbatim}
and so on. Full example \bytecodes{} are part of the repository as well as a
\lang{Python} script to dynamically construct such \bytecode{} for any $N$.
Building and running the code is explained in the \code{README}.

\subsection*{References}

\bibliographystyle{elsarticle-num}
\bibliography{vm_paper}

\end{document}